\documentclass[12pt, letterpaper]{article}
\usepackage{graphicx}
\usepackage{amsfonts}
\usepackage{amsmath}
\usepackage{amssymb}
\usepackage{amsthm}
\usepackage{bm}
\usepackage{bbm}
\usepackage{booktabs}
\usepackage{natbib}
\usepackage{url} 
\usepackage{appendix}
\usepackage{xcolor}
\usepackage{authblk}
\usepackage{geometry}
\usepackage{float}
\usepackage{hyperref}
\DeclareMathOperator{\E}{\mathbb{E}}
\DeclareMathOperator{\Var}{\text{Var}}
\DeclareMathOperator{\Cov}{\text{Cov}}

\newtheorem{prop}{ \noindent P{\footnotesize ROPOSITION}}[section]
\newtheorem{lemma}{ \noindent L{\footnotesize EMMA}}[section]

\newtheorem{remark}{ \noindent R{\footnotesize EMARK}}[section]

\title{\textbf{Towards Fair Predictions: Group Conditional Concordance Index to Quantify Fairness in Time-to-Event Prognostication}}

\author[1]{Haoyuan Wang, BS}
\author[2]{Riddhiman Bhattacharya, PhD}
\author[1]{Richardo Henao, PhD}
\author[3]{Daniel Wojdyla, MSc}
\author[1,3]{Chuan Hong, PhD}
\author[1,2]{\textsuperscript{*}Matthew Engelhard, MD, PhD}

\affil[1]{Department of Biostatistics and Bioinformatics, Duke University School of Medicine}
\affil[2]{ Duke AI Health}
\affil[3]{Duke Clinical Research Institute}
\affil[*]{m.engelhard@duke.edu}

\begin{document}

\maketitle

\begin{abstract}
Fairness metrics are essential for rigorously defining, quantifying, and mitigating biases in predictive models. While most existing metrics focus on binary classification tasks, fairness in time-to-event analyses has received limited attention. To address this gap, we propose a novel group fairness metric, the group-conditional Concordance Index (xCI), which extends Harrell’s Concordance Index (CI) by conditioning on group membership. The xCI measures both within-group and cross-group ranking accuracy in the presence of right-censored data. We formally define the xCI, prove that CI is a weighted average of xCIs across all possible group pairs, and develop a consistent estimator using inverse probability of censoring weights (IPCW). We further investigate the relationship between xCI and predicted risk scores through analytical derivations and simulation studies. To demonstrate its practical utility, we present two case studies: (i) assessing the fairness of survival models trained on harmonized data from the Framingham Offspring, MESA, and ARIC studies, and (ii) evaluating fairness in existing cardiovascular disease (CVD) risk prediction models using Truveta, a large-scale electronic health record (EHR) database. Our results show that xCI effectively detects biases across demographic groups that are overlooked by existing metrics. Overall, xCI provides a valuable tool for fairness assessment in survival analysis, particularly in constrained resource allocation settings, and complements existing fairness evaluation approaches.
\end{abstract}

\section{Introduction}
Predictive models are commonly used in healthcare settings to support clinical decision-making and provide prognostic information consistent with available evidence. When used as clinical decision-support, these models can significantly impact the allocation of limited resources such as preventive care among individual patients and between different patient subpopulations. For example, the Kidney Failure Risk Equation (KFRE) is used to prioritize referrals to nephrology care by estimating a patient’s risk of progression to kidney failure, thereby guiding early interventions for those at highest risk \citep{tangri2011predictive}. Similarly, risk prediction algorithms such as the CHA(2)DS(2)-VASc score help determine the need for anticoagulation in patients with atrial fibrillation to prevent stroke, ensuring that treatment is targeted to those who would benefit most \citep{lip2010refining}. Given the high impact of these decisions on patient health, it is critical to ensure that model-based resource allocation is fair to all groups. Here, we use “fair” in the algorithmic sense: individuals with comparable underlying clinical need should have comparable chances of being correctly prioritized by the model, regardless of demographic group membership.

However, in practice, biases in the underlying data and flaws in algorithm design can lead to systematically inaccurate or unfair predictions, especially for disadvantaged groups. For instance,  underrepresentation of certain demographic groups in datasets can lead to selection bias and skewed risk assessments~\citep{chen2021algorithm}. Similarly, Obermeyer et al. demonstrated that a widely adopted algorithm that uses health cost as a proxy for health needs will systematically underpredict the risk for African Americans~\citep{obermeyer2019dissecting}. When such a biased model is used to allocate clinical resources, such as referrals to specialized care or access to disease management programs, it can systematically divert resources away from the groups that need them most, thereby reinforcing existing disparities. Evaluating fairness of algorithms is the crucial first step in ensuring the transparency and accountability of AI-driven clinical decision-making system. Without identifying unfairness, interventions cannot be made, and healthcare disparities will exacerbate~\citep{nordling2019fairer} \citep{cary2025empowering} \citep{economou2024translating}.  However, while the preceding examples highlight clear violations of algorithmic fairness leading to unfair resource allocation and associated harms, in general defining fairness precisely can be challenging and controversial.

Ultimately, precise and statistical definitions of fairness are required to quantitatively evaluate predictive models and determine whether corrective actions are warranted. Many fairness concepts draw on philosophical foundations attributed to John Rawls's \textit{A Theory of Justice}, which provides a normative framework for thinking about fairness and equity in society. Rawls's Fair Equality of Opportunity principle asserts that individuals should have equal chances to succeed regardless of their background, while the Difference Principle holds that inequalities are only justifiable if they benefit the least advantaged members of society~\citep{rawls1971atheory}. In the context of healthcare, these principles motivate concern both for equal access to health-improving resources and for prioritizing those with the greatest medical need. 

We aim here to introduce a fairness metric that enables empirical evaluation of whether model-driven resource allocation aligns with these principles. Specifically, it allows us to assess whether individuals with similar clinical need have a similar chance to be correctly prioritized regardless of their demographic attributes (relating to the Fair Equality of Opportunity principle), and whether those with the most severe conditions are appropriately prioritized (relating to the Difference Principle). In this way, our framework provides a tool to critically examine the fairness of predictive models in clinical settings, informed by well-established ethical theory.

Building on this ethical foundation, fairness metrics provide practical tools to evaluate whether predictive models meet these principles in practice. Broadly, fairness metrics fall into two categories: (1) individual fairness metrics, which assess whether similar individuals receive similar predictions, and (2) group fairness metrics, which evaluate disparities in model performance or outcomes across demographic groups. Within group fairness, three key statistical notions have been widely studied: independence, separation, and sufficiency. These notions have been shown to be mathematically incompatible in most real-world settings, requiring modelers to prioritize one fairness criterion over the others depending on the context.~\citep{kleinberg2016inherent}

In the case of clinical risk prediction used for resource allocation, we focus on separation, because this notion aligns with Rawls's Fair Equality of Opportunity principle by ensuring that individuals with similar medical needs receive a comparable chance of being correctly identified by the model, regardless of demographic background. It also supports the Difference Principle by enabling more accurate identification of high-risk individuals who may benefit most from intervention. Compared to independence or sufficiency, separation is more efficient and actionable in clinical setting. 

Currently, most separation-based fairness metrics are designed for binary classification settings, with limited development for time-to-event outcomes~\citep{do2023fair}. For example, the xAUC metric has been proposed to evaluate fairness in resource allocation by assessing how well a model ranks individuals from different groups in a classification setting~\citep{kallus2019fairness}. While binary classification metrics like the AUC and xAUC can be adapted to time-to-event settings by evaluating predictions over a fixed time interval, this approach introduces sensitivity to the chosen time horizon, which must be selected with care~\citep{heagerty2000time}. 

Another approach involves comparing traditional survival analysis metrics (e.g., Concordance Index (CI)) across groups to assess fairness~\citep{sonabend2022flexible}. However, such comparisons primarily capture within-group fairness, which is how well the model performs within each group independently. While useful, this perspective misses between-group fairness, which evaluates how individuals from different groups are ranked relative to each other. between-group fairness is essential in contexts involving resource allocation, where decisions are inherently comparative across groups. Without explicitly assessing between-group ranking, a model may appear fair within groups but still systematically favor one group over another in practical decision-making. 
    
In this paper, we propose a group fairness metric for time-to-event settings that aligns with both Rawlsian principles of justice and the formal notion of separation. Our approach aligns with separation because it evaluates fairness conditional on the true outcome, focusing on how predicted risks align with actual event times. It reflects Rawlsian fairness by emphasizing equity in the prioritization of individuals across social groups—ensuring that the least advantaged are not systematically deprioritized in resource allocation decisions. Our proposed metric can be viewed as a group-conditional decomposition of the CI, where we condition on the group membership of both individuals in each pairwise comparison. This framing maintains robustness to censoring and enables a natural interpretation in right-censored survival data: it estimates the probability that the model correctly orders risk between two individuals, either from the same group or from different groups, based on observed event times. While similar concept of group-conditioned CI has been introduced as concordance imparity—defined as the maximum difference between group-specific concordance indices~\citep{9679035}—our contribution lies in rigorously analyzing the statistical properties of the metric and demonstrating its application. Specifically, we establish its relationship with CI, prove its statistical consistency, characterize its relationship to
predicted risk scores, and show how it can detect between-group unfairness in both synthetic and real-world medical datasets.

\section{Preliminaries}
We consider a binary classification task with \( n \) individuals, indexed by \( i = 1, \ldots, n \). Each individual \( i \) has a corresponding feature vector \( \mathbf{X}_i \in \mathcal{X} \), which may include various covariates and a \emph{group indicator} \( G_i \in \mathcal{G} \). We denote the true class label for individual \( i \) by \( Y_i \in \mathcal{Y} \). A trained model \( m: \mathcal{X} \to \mathcal{Y} \) assigns to each individual \( i \) a \emph{risk score} \( R_i \), which can be further thresholded to produce a binary prediction.

In survival analysis, the \emph{outcome} may not be observed for all individuals. Instead of a definite label such as \( 0 \) or \( 1 \), we often have a \emph{time-to-event} outcome. Let \( T_i \) be the (possibly unobserved) time at which a certain event of interest (e.g., death, failure, or relapse) occurs for individual \( i \). For each individual, there is also a \emph{censoring time} \( U_i \). If the event does not occur by the time one stops observing the individual, then \( U_i \) is smaller than \( T_i \). Consequently, the observed time for individual \( i \) is 
\( O_i = \min(T_i, U_i).\)
When \( O_i = T_i \), we say that the event was observed for individual \( i \), and otherwise, if \( O_i = U_i < T_i \), the individual is \emph{censored}.

Although in classical binary classification we directly observe \( Y_i \in \{0,1\} \), in survival analysis we typically define an \emph{event indicator} \( Y_i \) (also denoted \( \delta_i \) in some texts) as

\begin{align*}
    Y_i =
\begin{cases}
1, & \text{if } T_i \le U_i \quad (\text{event observed}), \\
0, & \text{if } T_i > U_i \quad (\text{censored}).
\end{cases}
\end{align*}

Thus, \( Y_i \) indicates whether the event was actually observed before censoring. Because we assume the event eventually occurs for every individual, \( T_i \) is finite but may lie beyond the censoring time \( U_i \). Consequently, for each individual we only observe the pair \( (O_i, Y_i) \)—the observed time \( \min(T_i, U_i) \) and whether or not the event happened during the observed period. These data \( \{(\mathbf{X}_i, O_i, Y_i)\}_{i=1}^n \) are then used to train a predictive model \( m(\cdot) \) that outputs a risk score \( R_i \). Depending on the analysis goal, one can convert \( R_i \) to a survival probability estimate or a hazard function estimate, or simply use a cutoff to yield a binary classification for downstream applications.

\section{Related Work}  

\subsection{Individual Fairness Metrics}
The individual fairness metrics focus on fairness at individual levels, assessing whether the model is making analogous predictions for similar individuals. For example, similarity-based metrics usually start with a clear definition of similarity, which measures how closely two individual relate based on their feature vectors, $\mathbf{X}_i$ and $\mathbf{X}_j$. The predictive model $m$ is fair if the difference of its prediction is small for a pair of similar individuals~\citep{dwork2011fairness}. For example, people often use the Lipschitz continuity condition to formalize the idea of individual fairness in the algorithm ~\citep{ilvento2020metric, pmlr-v119-mukherjee20a}

\begin{align*}
    d_\mathcal{Y}\big(m(\mathbf{X}_i), m(\mathbf{X}_j)\big) \le L d_\mathcal{X}(\mathbf{X}_i, \mathbf{X}_j) \quad \forall \mathbf{X}_i, \mathbf{X}_j \in \mathcal{X}
\end{align*}

 where $d_{\mathcal X}$ and $d_{\mathcal Y}$ are user-specified metrics on $\mathcal X$ and $\mathcal Y$, respectively, and $L$ is a Lipschitz constant for $m:\mathcal X\to\mathcal Y$ with respect to $(d_{\mathcal X},d_{\mathcal Y})$.The condition states that, for every pair of individuals, the difference in the prediction they receive from a model $m$ should be bounded by a Lipschitz constant times their distance in attributes~\citep{dwork2011fairness}. This condition ensures that smaller differences between individuals lead to small differences in predictions. 

However, similarity in attributes requires a context-based definition by humans rather than simply calculating the difference between a pair of individual attributes. For example, a young, obese person is different from an old, fit person, but they might have similar attributes in terms of developing stroke. Thus, different evaluations of similarity might introduce bias and human-in-the-loop error~\citep{fleisher2021s}.

\subsection{Group Fairness Metrics}

Group metrics, on the other hand, assume individuals are isolated and measure the aggregated prediction outcomes for different groups ~\citep{mehrotra2022revisiting}. Generally, group fairness metrics are based on three fairness criterion: Independence, Separation, Sufficiency, and these concept are incompatible with each other.

\subsubsection{Independence}
Independence asserts that predictions should not depend on group status~\citep{raz2021group}. For example, demographic parity reflect this idea, and ensures the proportion of people in each group receiving positive prediction is similar, i.e., 

\begin{align*}
    \Pr(m(\mathbf{X}_i) | G_i = a) = \Pr(m(\mathbf{X}_j) | G_i = b).
\end{align*}

 Resource allocation following independence has a straight forward intuition and could mitigate historical bias in the data. However, this idea might be inefficient if risk among different groups are disproportionate, which is common in the clinical setting~\citep{yao2017beyond}.
 
\subsubsection{Separation}
Separation allows heterogeneous risk among groups by ensuring that predictions should not depend on group status after conditioning on the outcome. For example, Equalized Odds state that a predictive algorithm is considered fair if it achieves equal False Positive Rate and False Negative Rate across different groups~\citep{NIPS2016_9d268236}, i.e.,

\begin{align*}
    \Pr(m(\mathbf{X}_i) = 0 | Y_i = 1, G_i = a) &= \Pr(m(\mathbf{X}_j) = 0 | Y_j = 1, G_i = b) \\
    \Pr(m(\mathbf{X}_i) = 1 | Y_i = 0, G_i = a) &= \Pr(m(\mathbf{X}_j) = 1 | Y_j = 0, G_i = b).
\end{align*}

This constraint ensures the predictive algorithm yields similar classification for people with the same true label from different groups. Another similar metric is balance for positive and negative classes~\citep{8452913}. This constraint states that after conditioning on the outcome, the predicted risk score for the two groups should be similar. 
\begin{align*}
    \E(R_i| Y_i = 1, G_i = a)  &=  \E(R_j| Y_j = 1, G_j = b) \\
     \E(R_i| Y_i = 0, G_i = a)  &=  \E(R_j| Y_j = 0, G_j = b).
\end{align*}

Recently, the notion of cross-group fairness has received increasing attention in the evaluation of predictive models. Traditional metrics such as the AUC primarily assess within-group performance—that is, how well a model distinguishes between positive and negative instances within the same demographic or subgroup. In contrast, the xAUC metric captures cross-group performance by measuring the probability that a positive instance from one group is ranked above a negative instance from a different group \citep{NEURIPS2019_73e0f748}. 

\begin{align*}
    \text{xAUC}_{a,b} = \Pr( R_i > R_j| Y_i = 1, Y_j = 0,  G_i = a, G_j = b ).
\end{align*}

This cross-group perspective provides a more comprehensive understanding of fairness, especially when assessing whether a model systematically favors one group over another in its rankings.

\subsubsection{Sufficiency}
Sufficiency emphasizes that prediction score should be equally informative about the true outcome across groups~\citep{kleinberg2016inherent}.

\begin{align*}
    \Pr(m(\mathbf{X}_i) = 1 | R_i = r, G_i = a) &= \Pr(m(\mathbf{X}_i) = 1 | R_j = r, G_j = b).
\end{align*}

Calibration, which states that the predicted probability of a group of people belonging to a class should match the proportion of people actually in that class, is closely related to the idea of sufficiency. A model is fair in terms of calibration, if the calibration within each group is similar. Calibration could be visualized using Calibration slope and intercept, where a slope close to 1 and intercept close to 0 indicates good calibration\citep{luo2022local}. Brier score also measures calibration to some extent, however, it has been shown that a lower Brier score does not necessarily indicate good calibration~\citep{rufibach2010use}.

\subsubsection{Incompatibility of Different Notions}
It is worth noting that these three notions are inherently incompatible. When the base rate of event occurring is not equal among groups, independence and separation, independence and sufficiency can not be achieved at the same time. Moreover, it has been shown that a model can not achieve perfect calibration and equal error rate or balance for positive and negative classes unless either it always predicts true risk, or the base rate is equal in both groups~\citep{kleinberg2016inherent, pleiss2017fairness}. In order to achieve equal error rate, a model might underpredict or overpredict the risk for one group, which means the predicted probabilities won’t match the actual outcomes as closely. As a result, a decision has to be made on what aspect of fairness should be preferred. In health resource allocation, independence it ensures health resource is divided among all the subgroup equally. When there is a significant event rate difference among subgroups, this will violates Rawlsian \textit{Difference principle}, since it is not benefiting people with worst condition.  Additionally, guaranteeing sufficiency is less robust compared to ensuring the order after conditioning on the true outcome. Since, order is more important in  constrained resource allocation, and perfect calibration could be achieved even if the order are totally incorrect. Therefore, we believe that our metric, which falls into the category of separation is well-suited to identify and quantify unfairness in health resource allocation guided by time-to-event predictive models, even though it does not explicitly guarantee independence and sufficiency.

\subsection{Metrics for Survival Analysis}
Most of the previous metrics focused on measuring the binary classification and could not be directly adopted to survival analysis due to censoring. Censoring is a common issue in survival analysis, and it refers to when only incomplete information is available for a subject. For example, a subject drops out of a study, but certain events free time is available. Excluding this information or treating this person event-free will lead to significant biases. Metric for survival analysis such as Concordance Index (CI) prevents this issue by incorporating information from limited observed time. 

The CI is defined as follows:

\begin{equation}\label{defn:CI}
    \text{CI} = \Pr(R_i > R_j \mid T_i < T_j).
\end{equation}

We estimate the CI by first identifying comparable pairs of individuals for which the observed data, specifically \(O_i\) and \(Y_i\), is sufficient to determine the order of the event occurring. Concordant pairs are defined as pairs whose model-predicted risk is consistent with the known event ordering. The estimator of CI is then calculated as the proportion of concordant pairs among all comparable pairs.

Researchers also adapt Area under the ROC Curve(AUC) to make it time-dependent and aggregate this metric over specific time intervals~\citep{kamarudin2017time, engelhard2025exploring}. However, this metric requires careful consideration of the time interval and has less straightforward interpretation. 

The metric proposed in this paper is based on the CI. CI evaluates the model's performance based on the pair-wise comparative risk of an event occurring, so it is robust to censoring. Our metric has adopted this characteristic CI and shows uncovered disparities in between-group fairness by conditioning on the group status. 

To our knowledge, the work most closely related to ours is the Concordance Imparity metric proposed by Zhang and Weiss~\citep{9679035}, which is defined as the maximum difference in group-conditional CI. However, our proposed metric focuses on the individual group-conditional CI itself. We further extend this line of work by rigorously analyzing its theoretical properties and demonstrating its utility in guiding fair allocation of healthcare resources.

\section {Group Conditional Concordance Index and Derived Measures of Fairness}\label{cross:concordance}
\subsection{Group Conditional Concordance Index (xCI)}

To introduce the concept of fairness to the concordance index as defined in \eqref{defn:CI}, we introduce the novel metric, asymmetric xCI, which provides insight into the model performance coupled with group fairness. In order to formally define the aforementioned metric, we condition on the group membership of both individuals in the calculation of the CI, which is given as

\begin{align}\label{defn:assym:xci}
    \text{xCI}_{a,b} = \Pr(R_i > R_j \mid T_i < T_j, G_i=a, G_j=b)
\end{align}

where $a,b \in \mathcal{G}$, with $\mathcal{G}$ being the set of all group labels. 

When choosing $a=b$, we obtain the within-group xCI, which is simply the CI limited to individuals from the chosen group. However, when choosing $a \ne b$, we have the between-group xCI. It compares performance of the model between groups, which is a crucial aspect for assessing fairness..

To estimate the xCI from data, we begin with the following key definitions. Denote the data and the risk prediction scores as $\{\bm{X}_i, O_i, Y_i\}_{i=1}^N$, and  $R_i$, respectively. Define
\begin{equation}\label{defn:comparable_pairs:xci}
\mathcal{S}_{a,b}
=
\{(i,j): i\neq j,\ O_i < O_j,\ Y_i=1,\ G_i=a,\ G_j=b\}.
\end{equation}
to be the set of all individuals from groups $a$ and $b$ for which we can determine the event ordering from data. We call these the \textit{comparable} pairs. We then define the \textit{concordant} pairs $\mathcal{S}_{a,b}^c$ as the subset of the comparable pairs $\mathcal{\mathcal{S}}_{a,b}$ whose predicted risk is consistent with the known event ordering:

\begin{equation*}\label{defn:concordant_pairs:xci}
    \mathcal{S}_{a,b}^c = \{(i, j) \in \mathcal{S}_{a,b} \mid  R_i > R_j\}
\end{equation*}

With these definitions in hand, we may estimate the xCI from data as the proportion of concordant pairs among the comparable pairs:

\begin{equation}\label{defn:estimator:xci}
    \hat{\text{xCI}}_{a,b} = \frac{|\mathcal{\mathcal{S}}_{a,b}^c|}{|\mathcal{\mathcal{S}}_{a,b}|}.
\end{equation}

Note that for any finite set $\mathcal{A}$, we write $|\mathcal{A}|$ to denote its cardinality, i.e., the number of elements in $\mathcal{A}$.

We now state a key result regarding xCI, relating it to CI,  which also elucidates why it is indeed important to define such a metric.

\paragraph{Assumption.}
Throughout, we assume the group label $G$ takes values in a finite set $\mathcal{G}$ and defines a partition of the population:
each individual belongs to exactly one group. Equivalently, the events $\{G=a\}_{a\in\mathcal{G}}$ are mutually exclusive and collectively exhaustive, i.e.,
$\Pr(G\in\mathcal{G})=1$ and $\Pr(G=a,\,G=b)=0$ for all $a\neq b$.
Under this assumption, the set of comparable pairs $\mathcal{S}$ admits the disjoint decomposition
\[
\mathcal{S} \;=\; \bigsqcup_{(a,b)\in\mathcal{G}^2}\mathcal{S}_{a,b},
\qquad
\mathcal{S}_{a,b} \;=\;\{(i,j)\in\mathcal{S}: G_i=a,\;G_j=b\},
\]
so that $\sum_{a,b\in\mathcal{G}}|\mathcal{S}_{a,b}|=|\mathcal{S}|$.

\begin{prop}\label{CI:weighted:prop}
Under above assumptions, the CI is a weighted average of xCIs.
\end{prop}
\begin{proof}
    The proof is deferred to the Appendix~\ref{appendix:weighted average}.
\end{proof}

\paragraph{Proof intuition.}
The CI is computed over the full set of comparable pairs $\mathcal S$,
whereas $\text{xCI}_{a,b}$ is the same concordance calculation restricted to the subset
$\mathcal S_{a,b}=\{(i,j)\in\mathcal S: G_i=a,\;G_j=b\}$.
Because the group labels form a partition of the population, the sets
$\{\mathcal S_{a,b}\}_{(a,b)\in\mathcal G^2}$ are disjoint and their union is $\mathcal S$.
Hence $|\mathcal S|=\sum_{a,b}|\mathcal S_{a,b}|$ and similarly
$|\mathcal S^c|=\sum_{a,b}|\mathcal S_{a,b}^c|$ for concordant pairs.
Within each block, $|\mathcal S_{a,b}^c|=|\mathcal S_{a,b}|\widehat{\text{xCI}}_{a,b}$ by definition,
so
\[
\widehat{\text{CI}}=\frac{|\mathcal S^c|}{|\mathcal S|}
=\sum_{a,b}\frac{|\mathcal S_{a,b}|}{|\mathcal S|}\,\widehat{\text{xCI}}_{a,b},
\]
a weighted average with weights $w_{a,b}=|\mathcal S_{a,b}|/|\mathcal S|$.
These weights depend on how frequently each type of group-pair comparable ordering occurs and  observed,
so directions with few comparable pairs can be down-weighted in the overall CI.

Since the CI is a weighted average of xCIs, it indicates that the CI is always between the xCIs, and the minimum xCI is always less than or equal to the CI. 

The weights represent the ratio of comparable pairs used to calculate a specific xCI to the overall comparable pairs, which are independent with prediction. This observation implies a potential limitation of the CI when there is a risk difference between groups. The CI depends on the true risk of the groups such that concordance between certain comparison pairs may be weighted minimally. For instance, if group $a$has a systematically higher survival time than group $b$, the cardinality of $\mathcal{S}_{a,b}$ will be much smaller than $\mathcal{S}_{b,a}$, resulting in lower weights for $\text{xCI}_{a,b}$. Consequently, it is possible to achieve a favorable CI even when the model performs poorly in correctly ranking individuals from group $a$ over $b$. 

\subsection{xCI-Based Fairness Diagnostics}

Fairness issues in pairwise risk ranking can arise through two qualitatively different mechanisms.
First, a model may separate risk more accurately within some groups than others, producing unequal within-group discrimination.
Second, even if within-group accuracy is similar, the model may behave asymmetrically in cross-group comparisons, correctly prioritizing one group ahead of another much more often than it correctly prioritizes the reverse ordering under the corresponding truth.
To distinguish these failure modes, we propose (i) disparity measures that separately quantify within-group gaps and directional cross-group asymmetry for a given pair of groups, and (ii) a worst-case summary across all ordered group pairs when multiple groups are present.

\subsubsection{xCI difference}
First, we define the within-group and between-group xCI disparity ($\Delta_{\text{within}} \text{xCI}_{a,b}$ and $\Delta_{\text{between}} \text{xCI}_{a,b}$) as follows to quantify the disparate effect between group $a$ and group $b$: 

\begin{equation*} \label{defn:within xCI diff}
    \Delta_{\text{within}} \text{xCI}_{a,b} = \text{xCI}_{a,a} - \text{xCI}_{b,b}.
\end{equation*}

\begin{equation*}\label{defn:between xCI diff}
    \Delta_{\text{between}} \text{xCI}_{a,b} = \text{xCI}_{a,b} - \text{xCI}_{b,a}.
\end{equation*}

These two values reveal different aspects of fairness. In $\Delta_{\text{within}} \text{xCI}_{a,b}$, two individuals that form comparable pairs come from the same group, and it can be viewed as the difference in accuracy of ranking within two groups. A large absolute value indicates the model is systematically more accurate in one group. 

On the other hand, $\Delta_{\text{between}} \text{xCI}_{a,b}$ illustrates how the model correctly ranks individuals from group $a$ ahead of individuals from group $b$ compared to the reverse ranking. A large absolute value is concerning because it indicates that the model effectively captures instances where the individual from group $a$ has a shorter event time than the individual from group $b$, but fails to do so in the opposite scenario. If we were to allocate preventive care using the model, individuals in group $a$ would have a systematically higher likelihood of receiving treatment based on correct rankings compared to those in group $b$.

Another way to understand the idea of between-group fairness and within-group fairness is by considering the scenario of deciding the treatment order for two patients based on predicted risk from a model in the Emergency Room(ER). If they belong to the same group, within-group fairness guarantees that prediction accuracy does not vary significantly no matter what group they both belong to. When they are from different groups, between-group fairness assures that the ranking accuracy is similar regardless of group membership of each patient. In an actual ER, we will have many patients from different groups that form both within-group and between-group comparison pairs. As a result, it is important to ensure both within-group and between-group fairness of the algorithm to adequately address all possible comparison scenarios.

It is important to note that within-group xCI can be obtained by calculating CI separately within each group, a method already used in several studies to assess fairness. However, Proposition~\ref{CI:weighted:prop} demonstrates that this approach fails to account for all possible comparison scenarios. Our xCI metric addresses this limitation by incorporating between-group comparisons, offering a more comprehensive assessment of fairness.

\subsubsection{Minimum xCI}
xCI differences become less straightforward when comparing more than two groups. For example, with four groups, there are 16 xCI metrics and 72 possible within- and between-group differences. Without a predefined focus on specific group pairs, a more general approach is to consider the minimum xCI ($\text{min xCI}$), defined as the smallest xCI value across all group pairs in $\mathcal{G}$:

\begin{equation*}\label{defn:min xCI }
    \text{min xCI} = \min_{a,b \in \mathcal{G}}\text{xCI}_{a,b}.
\end{equation*}

A large value is preferred for $\text{min xCI}$ since it represents the model's worst performance among all possible within-group and between-group comparisons. 

These simple checks could also be used as loss functions in machine learning algorithms aiming for fairness. For example, we could minimize the xCI difference or maximize the min xCI. However, these simple approaches have trade-offs in terms of accuracy: Algorithms that minimize xCI differences may encourage lowering the performance for comparable pairs where the algorithm works well. On the other hand, maximizing the min xCI might overlook solutions that could significantly improve overall performance by slightly hurting the accuracy in worst the comparable pairs. 

\subsection{xCI based Utility}
A dichotomy exists between the broader philosophical and ethical dimensions of fairness and the quantifiable measures provided by fairness metrics \citep{binns2018fairness}; while fairness captures overarching societal and moral considerations, fairness metrics focus on specific, measurable aspects of model behavior. Extra modification and justification should be considered to use metrics to measure fairness.

In a clinical setting, both "correctly ranking before" and "correctly ranking after" can benefit patients, depending on the context. For instance, consider a colon cancer prediction model designed to inform decisions about colonoscopy treatment for patients from two groups, a and b. Correctly ranking group $a$ before group $b$ not only enhances recovery outcomes for patients in group $a$ but also spares patients in group $b$ from unnecessary and potentially harmful interventions. Thus, $\text{xCI}_{a,b}$ quantifies the model’s ability to correctly rank individuals from group $a$ ahead of those from group $b$ (benefiting group $a$) while simultaneously ranking individuals from group $b$ behind those from group $a$ (benefiting group $b$).

Given these considerations, the xCI metric can be weighted to reflect the comparative utility that one group gains from the predictive algorithm. Specifically, $\text{xCI}_{a,b}$ and $\text{xCI}_{b,a}$ measure the model's accuracy in ranking patients under two distinct scenarios. The utility a patient derives from the prediction algorithm can then be expressed as follows:

\begin{equation*}\label{defn: expectation xCI}
    U(a) = \alpha \text{xCI}_{a,b}+ \beta \text{xCI}_{b,a}.
\end{equation*}

To extend this to multi group comparison:
\begin{equation*}\label{defn: expectation group a}
    U(a) = \alpha \sum_{b \in \mathcal{G}, b \neq a}\text{xCI}_{a,b}+ \beta \sum_{b \in \mathcal{G}, b \neq a}\text{xCI}_{b,a}
\end{equation*}

where 
\begin{equation*}\label{defn: xCI a and all other group}
   \sum_{b \in \mathcal{G}, b \neq a}\text{xCI}_{a,b} = Pr( R_a > R_{a'} | T_a < T_{a'}), 
\end{equation*}

and $a'$ represent the union of all groups $b \ne a$.

Once the utility is obtained, we could use fair utility optimization techniques such as maximizing the minimum utility for different groups. ~\citep{yang2023minimax} Alternatively, we could apply the maximum Nash welfare solution, which maximizes the product of utilities across different groups~\citep{caragiannis2019unreasonable}. 

\subsection{IPCW estimator for $\mathrm{xCI}$ under group-specific censoring}
The naïve estimator in \ref{defn:estimator:xci} can be biased under right censoring because the set of pairs for which the earlier time is observed is, in general, not a random sample of all comparable pairs. This bias may be corrected through the inverse probability of censoring weighting (IPCW) as described by~\citep{uno2011c}.

Let $\Delta_i := I(T_i \le U_i)$ be the event indicator, and $R_i$ a risk score. For any two groups $a,b\in\mathcal G$ and a truncation time $\tau<\infty$, define the truncated cross-group concordance
\begin{equation}
\label{eq:xci_target}
\mathrm{xCI}_{a,b,\tau}(R)
:= \Pr\!\left(R_i > R_j \,\middle|\, T_i < T_j,\; T_i < \tau,\; G_i=a,\; G_j=b\right).
\end{equation}
We truncate at $\tau$ to avoid instability in the tail of the censoring survival function and assume $\Pr(U>\tau\mid G=g)>0$ for the groups of interest~\citep{heagerty2005survival}.

Let the group-specific censoring survival functions be
\[
K_g(t) := \Pr(U>t \mid G=g), \qquad g\in\mathcal G,
\]
and let $\widehat K_g(t)$ be the Kaplan--Meier (KM) estimator of $K_g(t)$ within group $g$, estimating the censoring distribution by treating censoring as the ``event'' (i.e., using $(O_i,\,1-\Delta_i)$ among individuals with $G_i=g$).
Define the group-specific IPCW weight for comparing groups $a$ and $b$ by
\begin{equation}
\label{eq:ipcw_weight}
\widehat w_{a,b}(t) := \widehat K_a(t)^{-1}\widehat K_b(t)^{-1}.
\end{equation}

Then the IPCW estimator of $\mathrm{xCI}_{a,b,\tau}(R)$ is
\begin{equation}
\label{eq:ipcw_xci_est}
\widehat{\mathrm{xCI}}_{a,b,\tau}
:=
\frac{
\sum_{i:\,G_i=a}\ \sum_{\substack{j:\,G_j=b\\ j\neq i}}
\Delta_i\, \widehat w_{a,b}(O_i)\, I(O_i<O_j,\; O_i<\tau)\, I(R_i>R_j)
}{
\sum_{i:\,G_i=a}\ \sum_{\substack{j:\,G_j=b\\ j\neq i}}
\Delta_i\, \widehat w_{a,b}(O_i)\, I(O_i<O_j,\; O_i<\tau)
}.
\end{equation}

\paragraph{Assumptions.}
Fix $\tau<\infty$.
\begin{enumerate}
\item \textbf{Sampling and group positivity.}
The observations $\{(T_i,U_i,G_i,R_i)\}_{i=1}^n$ are i.i.d., and
$\Pr(G=a)>0$ and $\Pr(G=b)>0$.

\item \textbf{Non-informative censoring within groups and weight positivity.}
Within each group, censoring is independent of event time and risk score:
\[
U \perp (T,R)\mid G,
\]
and the censoring survival is bounded away from zero on $[0,\tau]$ for the groups compared:
\[
\inf_{t\in[0,\tau]} K_g(t) \ge \kappa > 0,\qquad g\in\{a,b\}.
\]

\item \textbf{Consistency of censoring weights.}
For $g\in\{a,b\}$, the group-specific KM estimator satisfies
\[
\sup_{t\in[0,\tau]} \bigl|\widehat K_g(t)-K_g(t)\bigr| \xrightarrow{P} 0.
\]
\end{enumerate}

\begin{prop}
\label{prop:ipcw_consistency_main}
Under Assumptions 1--3, the IPCW estimator in \eqref{eq:ipcw_xci_est} is consistent:
\[
\widehat{\mathrm{xCI}}_{a,b,\tau} \xrightarrow{P} \mathrm{xCI}_{a,b,\tau}(R),
\]
where $\mathrm{xCI}_{a,b,\tau}(R)$ is defined in \eqref{eq:xci_target}.
\end{prop}

\begin{proof}
The proof is deferred to Appendix~\ref{app:ipcw_proof}.
\end{proof}

\paragraph{Proof intuition.}
Under non-informative censoring within groups, a pair with $G_i=a,G_j=b$ and event time $T_i$ is observed only if $U_i\ge T_i$ and $U_j>T_i$, which occurs with probability $K_a(T_i)K_b(T_i)$.
IPCW multiplies each observed pair by $\{K_a(T_i)K_b(T_i)\}^{-1}$, so in expectation the censoring probabilities cancel, yielding the same numerator/denominator as if censoring were absent.
Replacing $K_g$ by $\widehat K_g$ does not change the limit because $\widehat K_g$ is uniformly consistent on $[0,\tau]$ and bounded away from zero.

\section{Impact of Predicted Group Differences on xCI}
\label{impact_predicted_group_diff}

In this section, we examine how differences in predicted risk between groups affect the cross-group concordance index $\text{xCI}_{a,b}$, where $a$ and $b$ are two distinct subgroups. We consider the We assume a log normal model with i.i.d.\ homoskedastic errors.

\begin{align} \label{log_normal_model}
    \log(T_i)=\beta_0+\beta_1 G_i+\beta_2 Z_i+\epsilon_i,
    \qquad \epsilon_i\sim\mathcal N(0,\sigma^2),
\end{align}

 where $G_i \;=\; I\{i \in a\}$, and $Z_i\sim\mathcal N(0,\sigma_z^2)$ is a continuous covariate, independent of $\epsilon_i$ and $G_i$.
We evaluate a linear risk score $\hat r_i=\hat\beta_1 G_i+\hat\beta_2 Z_i$, because the intercept is insensitive to ranks.

 For pairs $(i,j)$ with $i\in a$ and $j\in b$, we condition on the true ordering of events $T_i<T_j$ and study the probability that the score ranks $i$ above $j$. Let $\Delta_{ij} Z:=Z_i-Z_j$ and denote by $(\mu_c,\sigma_c^2)$ the conditional mean and variance of $\Delta_{ij} Z$ given $T_i<T_j$ (equivalently, given $D<0$ for $D:=\beta_1+\beta_2\Delta_{ij} Z-\Delta\epsilon$, see Appendix~\ref{appendix:impact_predicted_group_diff}).

\paragraph{Assumptions.}
Fix $\tau<\infty$.
\begin{enumerate}
\item \textbf{I.i.d.\ sampling across individuals.}
The observations $\{(T_i,G_i,Z_i,\epsilon_i)\}_{i=1}^n$ are i.i.d.

\item \textbf{Distribution and independence of the covariate $Z$.}
The covariate is Gaussian and independent of group membership and the error term:
\[
Z_i\sim \mathcal N(0,\sigma_z^2),\qquad Z_i \perp G_i,\qquad Z_i \perp \epsilon_i.
\]

\item \textbf{Pairwise comparability and no ties.}
For two independent draws $(i,j)$, ties occur with probability zero:
\[
\Pr(T_i=T_j)=0,
\]
so the ordering event $T_i<T_j$ is well-defined.
\end{enumerate}

\begin{prop}\label{prop:impact_predicted_group_diff}
Under Assumptions 1-3 model as in~\ref{log_normal_model}, for cross-group pairs $(i\in a,j\in b)$, we have
\[
\text{xCI}_{a,b}
=\Pr\!\big(\hat r_i>\hat r_j\mid T_i<T_j\big)
=\Phi\!\left(
\frac{\hat\beta_1+\hat\beta_2\,\mu_c}{\sqrt{\hat\beta_2^2\,\sigma_c^2}}
\right),
\]
where $\Phi$ is the standard normal Cumulative Density Function (CDF) and $(\mu_c,\sigma_c^2)$ depends only on $(\beta_1,\beta_2,\sigma^2,\sigma_z^2)$ via a truncated-normal adjustment.
\end{prop}

\begin{proof}
    The proof is deferred to the Appendix~\ref{appendix:impact_predicted_group_diff}.
\end{proof}

\begin{remark}
By symmetry, 
\[
\text{xCI}_{b,a}
=\Pr\!\big(\hat r_i>\hat r_j\mid T_i<T_j\big)
=\Phi\!\left(
-\frac{\hat\beta_1+\hat\beta_2\,\mu_c'}{\sqrt{\hat\beta_2^2\,\sigma_c'^2}}
\right),
\]
where $(\mu_c',\sigma_c'^2)$ are the corresponding conditional moments of $\Delta_{ij} Z$ given $T_i<T_j$ in the $b$ vs.\ $a$ pairing.

For $(i,j)$ in the same group, the group shift cancels so $\hat r_i-\hat r_j=\hat\beta_2\Delta_{ij} Z$, and
\[
\text{xCI}_{a,a}=\text{xCI}_{b,b}
=\Phi\!\left(\frac{\hat\beta_2\,\mu_{w}}{\sqrt{\hat\beta_2^2\,\sigma_{w}^2}}\right),
\]

with $(\mu_{w},\sigma_{w}^2)$ the conditional moments of $\Delta_{ij} Z$ given $T_i<T_j$ when $G_i=G_j$. The proof is deferred to Appendix~\ref{appendix:impact_predicted_group_diff}, where detailed information on  $(\mu_c,\sigma_c^2, \mu_w,\sigma_w^2)$ can be found. The equality $\text{xCI}_{a,a}=\text{xCI}_{b,b}$ is not guaranteed in general; we assume it here only for simplicity. In reality, each group may have a distinct within-group covariate and event-time distribution, so the corresponding conditional moments $(\mu_w,\sigma_w^2)$ can differ across $a$ and $b$.

\end{remark}
\paragraph{Proof intuition.}
For a cross-group pair $(i\in a,\;j\in b)$, the true ordering $T_i<T_j$ can be rewritten as a linear inequality in the Gaussian latent variables:
\[
T_i<T_j
\quad\Longleftrightarrow\quad
\log T_i-\log T_j <0
\quad\Longleftrightarrow\quad
D:=\beta_1+\beta_2(Z_i-Z_j)-(\epsilon_j-\epsilon_i) < 0.
\]
Since $\Delta Z:=Z_i-Z_j$ and $\Delta\epsilon:=\epsilon_j-\epsilon_i$ are independent Gaussians, $(\Delta Z, D)$ is jointly normal. Conditioning on the event $D<0$ therefore amounts to a \emph{truncation} of a normal variable, so the conditional distribution of $\Delta Z\mid(D<0)$ remains normal with a shifted mean $\mu_c$ and variance $\sigma_c^2$ (given by standard truncated-normal moment formulas).
Because the score difference is affine in $\Delta Z$,
\[
\Delta \hat r:=\hat r_i-\hat r_j=\hat\beta_1+\hat\beta_2\,\Delta Z,
\]
it follows that $\Delta\hat r\mid(D<0)$ is also normal with mean $\hat\beta_1+\hat\beta_2\mu_c$ and variance $\hat\beta_2^2\sigma_c^2$. Hence the cross-group concordance is just the probability a normal variable is positive:
\[
\text{xCI}_{a,b}=\Pr(\Delta\hat r>0\mid D<0)
=\Phi\!\left(\frac{\hat\beta_1+\hat\beta_2\mu_c}{\sqrt{\hat\beta_2^2\sigma_c^2}}\right).
\]
The reverse direction ($b$ vs.\ $a$) corresponds to conditioning on $D>0$ instead, producing the analogous expression with right-truncated moments. Within-group, the group shift cancels ($\Delta\hat r=\hat\beta_2\Delta Z$), so $\hat\beta_1$ cannot affect within-group discrimination.

Holding the truncation-induced moments $(\mu_c,\sigma_c)$ fixed for interpretation, Proposition~\ref{prop:impact_predicted_group_diff} implies that increasing the fitted group shift $\hat\beta_1$ raises $\text{xCI}_{a,b}$
thereby improving cross-group ranking when group $a$ truly tends to fail earlier than $b$, while, by symmetry, lowering $\text{xCI}_{b,a}$. Thus, our metric makes the directional trade-off explicit: improving performance in the $a$-over-$b$ comparison by elevating group $a$’s predicted risk necessarily degrades performance in the reverse direction. By contrast, within-group discrimination is invariant to $\hat\beta_1$ because the group shift cancels, so traditional within-group $\text{xCI}$ is blind to this effect. Recognizing this trade-off is crucial for fairness-aware model design, where one aims to balance performance across both within-group and between-group scenarios.

\subsection{Simulation Study}

To validate the above derived expression for $\text{xCI}$ derived in the previous section, we conducted a simulation study. Specifically, we investigated how the xCI metric varies as a function of the estimated group effect $\hat{\beta}_1$, holding all other parameters fixed.

We simulated survival times under \ref{log_normal_model}, where $G_i \in \{0, 1\}$ indicates group membership, and $Z_i \sim \mathcal{N}(0, \sigma_z^2)$ is a continuous covariate. The true parameter values were fixed at:
\[
\beta_1 = 0.8, \quad \beta_2 = 1.0, \quad \sigma_z = 0.5, \quad \sigma = 0.5.
\]

For each simulation run, we generated $n = 500$ individuals with equal probability of belonging to either group ($P(G=1) = 0.5$). The survival time $T_i$ was obtained by exponentiating the linear predictor. Risk scores were computed using a perturbed coefficient $\hat{\beta}_1 \in [0.2, 1.2]$ while fixing $\hat{\beta}_2 = 1.0$.

To compute the empirical $\text{xCI}_{a,b}$, we selected all comparable pairs $(i, j)$ such that $G_i = 1$, $G_j = 0$, and evaluated the proportion of times the risk score correctly ranked the individual with the shorter survival time. The theoretical xCI values were computed using the analytical formula derived from the theory section.

Each scenario (i.e., each value of $\hat{\beta}_1$) was repeated over $n_\text{sim} = 50$ independent simulations. The mean and standard deviation of the empirical xCI values were recorded.

\begin{figure}
    \centering
    \includegraphics[width=0.75\linewidth]{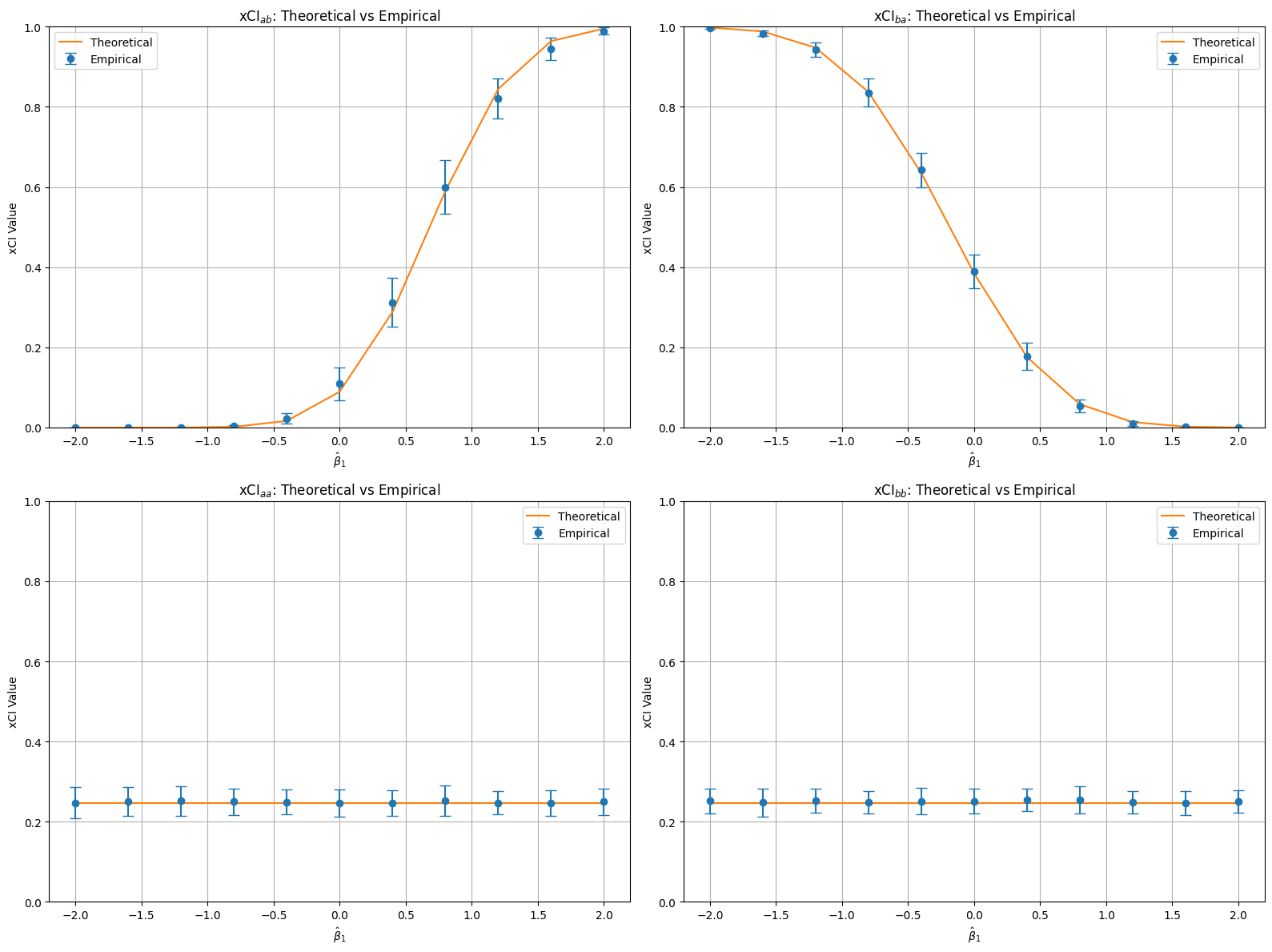}
    \label{fig:xci_simulation}
    \caption{Theoretical versus empirical xCI. Solid lines show the closed-form xCI values derived under the assumed data-generating model, while points show Monte Carlo estimates computed from simulated datasets. The error bars indicate Monte Carlo uncertainty Panels correspond to the four xCI conditioning configurations shown in the titles.}

\end{figure}

Figure~\ref{fig:xci_simulation} compares the empirical and theoretical xCI values across varying $\hat{\beta}_1$. The empirical results closely follow the theoretical curve, validating the derived expression.

\section {Motivating Use of xCI}

In this section, we demonstrate how xCI can reveal disparities in survival prediction models that are not captured by Harrell's CI. Using the flchain dataset, we compare several Cox proportional hazards models, including penalized variants and a model excluding sex as a predictor, to evaluate fairness with respect to sex. We show that between-group xCIs can uncover ranking disparities disadvantaging female participants, even when the sensitive attribute is omitted from the model. Moreover, to illustrate a setting where sex is indirectly encoded through other predictors, we synthetically induce correlation between sex and serum free light-chain measurements by shifting kappa (serum free immunoglobulin light chain, $\kappa$ portion) upward by 1 and lambda (serum free immunoglobulin light chain, $\lambda$ portion) downward by 1 for all female participants. This perturbation highlights the limitations of fairness through unawareness, and illustrates how xCI can diagnose resulting cross-group ranking disparities.

We evaluate the performance of xCI in the flchain data set~\citep{dispenzieri2012use}. This publicly available data aims to investigate the relationship between serum-free light chain (FLC) and mortality. The dataset contains demographic and clinical information for 7874 subjects sampled from residents aged 50 or greater of Olmsted County, Minnesota. We constructed a Cox Proportional Hazard (Cox PH) and Cox PH with Ridge Penalizer (Ridge), Cox PH with Lasso Penalizer (Lasso), and Cox PH without sex variable to predict mortality. The group-sensitive attribute we are interested in is sex, a binary variable with 1 indicating female and 0 otherwise. Note that the model constructed without sex variable is consist with the concept of fairness through unawareness, which is a fairness principle in machine learning that suggests an algorithm should be fair if it does not explicitly use sensitive attributes~\citep{10.1145/2090236.2090255}.



Then, we calculated IPCW xCI for female and male across four models, and the result is shown in Table~\ref{compelling example: xCI result table}. We observe that apart from without sex model, all other models have more significant differences between between-group xCI than within-group xCI, which implies unfairness in ranking that traditional CI metric fails to capture.  The difference in between-group xCI shows that all four models are more likely to correctly rank a male before a female than vice versa. If either of these prediction models were to inform the distribution of preventive care, females would be a disadvantaged group since our model tends to rank them incorrectly after males. 

\begin{table}[H]
\centering
\begin{tabular}{lcccccc}
\toprule
Model & $xCI_{male,male}$ & $xCI_{female,female}$ & $xCI_{male,female}$ & $xCI_{female,male}$ & $\Delta_{\text{within}}$ & $\Delta_{\text{between}}$ \\
\midrule
\textbf{Before Change} & & & & & & \\
Cox PH       & 0.5980 & 0.6678 & 0.8046 & \textbf{0.4264} & -0.0698 & 0.3782 \\
Ridge        & 0.5980 & 0.6678 & 0.8044 & \textbf{0.4266} & -0.0698 & 0.3778 \\
Lasso        & 0.5981 & 0.6681 & 0.7561 & \textbf{0.4926} & -0.0700 & 0.2635 \\
Without Sex  & 0.6005 & 0.6688 & 0.6405 & \textbf{0.6275} & -0.0683 & 0.0130 \\
\textbf{After Change} & & & & & & \\
Cox PH       & 0.5980 & 0.6678 & 0.8046 & \textbf{0.4264} & -0.0698 & 0.3782 \\
Ridge        & 0.5980 & 0.6677 & 0.8045 & \textbf{0.4265} & -0.0697 & 0.3780 \\
Lasso        & 0.5965 & 0.6671 & 0.7542 & \textbf{0.4928} & -0.0706 & 0.2614 \\
Without Sex  & 0.5960 & 0.6661 & 0.7749 & \textbf{0.4646} & -0.0701 & 0.3103 \\
\bottomrule
\end{tabular}
\caption{Cross-group concordance index (xCI) values before and after the change, including within-group difference $\Delta_{\text{within}} = \text{xCI}_{male,male} - \text{xCI}_{female,female}$ and between-group difference $\Delta_{\text{between}} = \text{xCI}_{male,female} - \text{xCI}_{female,male}$. Minimum value in each row is highlighted in \textbf{bold}. Change refers to adding correlation between sex and covariates.} 
\label{compelling example: xCI result table}
\end{table}

\begin{figure}[H]
    \centering
    \includegraphics[width=1\linewidth]{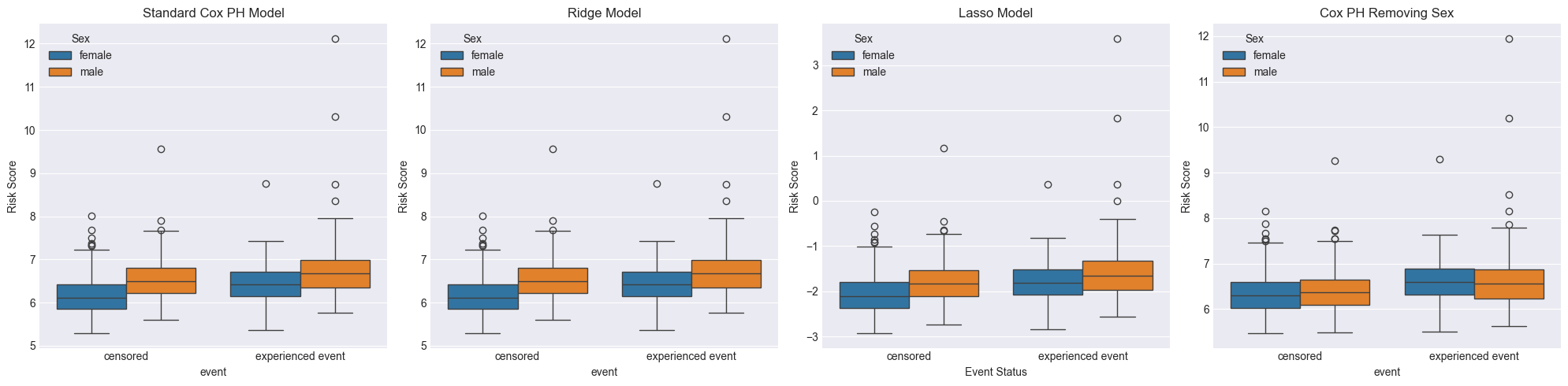}
    \caption{Distribution of predicted risk scores by sex and event status before introducing a correlation between sex and other covariates. The model excluding sex shows minimal between-group disparity in predicted risk scores, consistent with the reduced difference in between-group xCIs.}
    \label{fig:Predicted Risk Score Before Adding Association}

    \vspace{1em} 

    \includegraphics[width=1\linewidth]{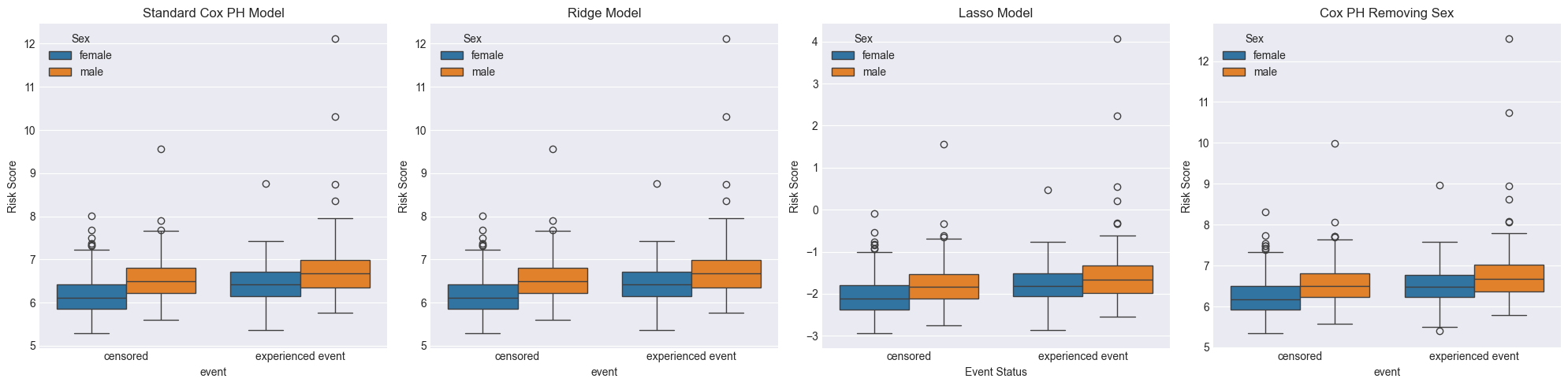}
    \caption{Distribution of predicted risk scores by sex and event status after increasing Kappa by one and decreasing Lambda by one for all female participants. The correlation between sex and covariates leads to larger disparities in predicted risk scores, particularly for the model excluding sex, which now shows the largest difference in between-group xCIs. This illustrates how fairness through unawareness fails when sensitive attributes are indirectly encoded through correlated covariates.}
    \label{fig:Predicted Risk Score After Adding Association}
\end{figure}

After removing the sex variable, we observed a significant reduction in disparity in xCIs. However, this effect diminishes when the sensitive attribute is strongly correlated with other covariates. To demonstrate this, we manually introduce an association by increasing Kappa by one and decreasing Lambda by one for all female participants in the dataset. In this modified dataset, the model without the sensitive attribute no longer shows the lowest disparity in xCI. Equipped with our proposed metric, we illustrate the limitation of fairness through unawareness: it fails achieve fair result when covariates are correlated with sensitive attributes.

Figures \ref{fig:Predicted Risk Score Before Adding Association} and \ref{fig:Predicted Risk Score After Adding Association} illustrate the distribution of predicted risk scores conditioned on the event and sex. Assuming non-informative censoring, a fair model that aligns with the notion of separation should yield predicted risk scores that are independent of group membership. From Table \ref{compelling example: xCI result table}, our metric effectively captures this concept. Models with large disparities in between-group xCI tend to exhibit greater divergence in the distribution of predicted risk scores. For example, before adding the association, the model without sex shows the smallest difference in xCI and similar distributions of predicted scores. However, after adding the association, the $\Delta_{\text{between}}\text{xCI}_{a,b}$ increases and becomes the largest among all the models, with the distributions of risk scores for males and females diverging the most.

These results supports the necessity of our metrics in spotting and quantifying the unfairness in the prediction model. When straightforward solutions like removing sensitive group variables prove ineffective, incorporating xCI into the machine learning algorithm as a penalty factor for unfairness can be a viable alternative.

\section {Case Study}
In this section, we present two case studies to illustrate how xCI can be applied and interpreted in real-world datasets. The first case study uses cohort data, where we construct local models to evaluate xCI. The second case study leverages Electronic Health Record (EHR) data to validate an existing risk prediction equation.

\subsection{Stroke Risk Prediction}
We conducted a study to predict 10-year stroke risk using harmonized data from three major epidemiological cohorts: the Framingham Offspring Cohort~\citep{kannel1979investigation}, the Multi-Ethnic Study of Atherosclerosis (MESA)~\citep{bild2002multi}, and the Atherosclerosis Risk in Communities (ARIC) study~\citep{aric1989atherosclerosis}. This harmonized dataset includes comprehensive demographic, clinical, and biomarker data for participants. We focus on Black and White participants to assess potential disparities in model performance.

We employed three common survival analysis models to estimate the 10-year stroke risk. The first model was the Cox PH model, which assumes a constant hazard ratio over time. The second model was the Random Survival Forest (RSF) ~\citep{ishwaran2008random}, a non-parametric ensemble learning method. The third model was the Accelerated Failure Time (AFT) model ~\citep{wei1992accelerated}, which directly models survival times and assumes that covariates act multiplicatively on the time scale.

To investigate potential disparities in prediction performance, we calculated the xCIs for black and white subgroups. 

\begin{table}[H]
\centering
\begin{tabular}{lcccccc}
\toprule
Model & $xCI_{black,black}$ & $xCI_{white,white}$ & $xCI_{black,white}$ & $xCI_{white,black}$ & $\Delta_{\text{within}}$ & $\Delta_{\text{between}}$ \\
\midrule
Cox PH & 0.7184 & 0.7543 & 0.8349 & \textbf{0.6218} & -0.0359 & 0.2131 \\
RSF    & 0.7053 & 0.7449 & 0.8290 & \textbf{0.6016} & -0.0396 & 0.2274 \\
AFT    & 0.6551 & 0.6694 & 0.7462 & \textbf{0.5702} & -0.0143 & 0.1760 \\
\bottomrule
\end{tabular}
\caption{Comparison of xCI values for different stroke risk prediction models, including within-group difference $\Delta_{\text{within}} = \text{xCI}_{black,black} - \text{xCI}_{white,white}$ and between-group difference $\Delta_{\text{between}} = \text{xCI}_{black,white} - \text{xCI}_{white,black}$. Minimum value in each row is highlighted in \textbf{bold}.}
\label{tab:xci_comparison}
\end{table}

From Table \ref{tab:xci_comparison}, we observe that all three models exhibit comparable within-group $\text{xCI}$ values, with $\text{xCI}_{white,white}$ slightly higher than $\text{xCI}_{black,black}$. This suggests that the models are marginally more accurate within the white subpopulation. In contrast, there is a notable disparity in the between-group $\text{xCI}$. Specifically, $\text{xCI}_{black,white}$ exceeds $\text{xCI}_{white,black}$, indicating that the models are more likely to correctly rank a black individual before a white individual than vice versa.

Among the models, the AFT model shows the poorest overall performance, particularly in cases where white individuals have shorter event times than black individuals. When applied to estimate stroke risk for a pair of black and white patients, the AFT model performs well when the black patient has a higher risk. However, when the white patient has a higher risk, the model's ranking accuracy is only slightly better than random guessing. This asymmetric performance indicates that the model's accuracy is highly dependent on which group has the higher risk, which is undesirable and may introduce disparities in downstream clinical decision-making.

\subsection{PREVENT Equation}
 In this case study, we evaluated the fairness of an existing clinical risk prediction algorithm for time-to-event outcomes using an external online EHR database. The Predicting Risk of CVD Events (PREVENT) model, developed by the American Heart Association (AHA), estimates the risk of total cardiovascular disease (CVD), which includes both atherosclerotic cardiovascular disease (ASCVD) and heart failure (HF)~\citep{Prevent}. The PREVENT model is sex-specific, calculating risk scores separately for males and females, and excludes race as a predictor. According to the developers, the model demonstrates strong discrimination and calibration.

For external validation, we used data from Truveta~\citep{TruvetaGuidelines}, a health data platform containing de-identified patient records from healthcare systems across the United States. Starting from more than 100 million individuals in the database, we derived a cohort of approximately 20,000 patients by applying the same inclusion and exclusion criteria used in the PREVENT study.

\begin{table}[H]
\centering
\begin{tabular}{lcccccc}
\toprule
\textbf{Equation} & $\text{xCI}_{male,male}$ & $\text{xCI}_{female,female}$ & $\text{xCI}_{male,female}$ & $\text{xCI}_{female,male}$ & $\Delta_{\text{within}}$ & $\Delta_{\text{between}}$ \\ 
\midrule
\textbf{CVD}    & 0.745 & 0.825 & 0.815 & \textbf{0.755} & -0.080 & 0.060 \\ 
\textbf{CVD*}   & 0.757 & \textbf{0.794} & NA    & NA    & -0.037 & NA \\ 
\textbf{ASCVD}  & 0.698 & 0.801 & 0.792 & \textbf{0.700} & -0.103 & 0.092 \\ 
\textbf{ASCVD*} & 0.736 & \textbf{0.774} & NA    & NA    & -0.038 & NA \\ 
\textbf{HF}     & 0.823 & 0.861 & 0.862 & \textbf{0.823} & -0.038 & 0.039 \\ 
\textbf{HF*}    & 0.809 & \textbf{0.830} & NA    & NA    & -0.021 & NA \\ 
\bottomrule
\end{tabular}
\caption{xCI values for male and female of PREVENT Equations (Base Model). ``*'' indicates the results from original paper~\citep{Prevent}. $\Delta_{\text{within}} = \text{xCI}_{male,male} - \text{xCI}_{female,female}$ and $\Delta_{\text{between}} = \text{xCI}_{male,female} - \text{xCI}_{female,male}$. Minimum value in each row is highlighted in \textbf{bold}.}
\label{PreventResult}
\end{table}



From Table \ref{PreventResult}, we observe that PREVENT exhibits a slightly larger $\Delta_{\text{within}}\text{xCI}_{\text{male},\text{female}}$ compared with the results reported in the original paper. This minor variation is expected, given that our evaluation is based on an external cohort. Consistent with the original findings, the CI for females is higher than that for males, suggesting that PREVENT achieves slightly better within-group discrimination for women.

However, examining $\text{xCI}{\text{male},\text{female}}$ and $\text{xCI}{\text{female},\text{male}}$ reveals an important nuance: males are more likely to be ranked correctly before females than the reverse. If PREVENT were used to allocate preventive care, this would imply that males are more likely to receive appropriate treatment before females, while females are more likely to be incorrectly prioritized after males.

This case study demonstrates how $\Delta_{\text{between}}\text{xCI}_{\text{male},\text{female}}$ provides additional insight into fairness concerns. Although the higher $\Delta_{\text{within}}\text{xCI}_{\text{male},\text{female}}$ suggests better within-group performance for females, the between-group ranking disparity indicates that women are not necessarily advantaged in the allocation process.

It is also notable that the between-group difference $\Delta_{\text{between}}\text{xCI}_{\text{male},\text{female}}$ is smaller than the within-group difference $\Delta_{\text{within}}\text{xCI}_{\text{male},\text{female}}$, in contrast to the pattern observed in the stroke example. This suggests that PREVENT does not exhibit substantial cross-group ranking disparities.

\section {Discussion}
We recognize that predictive models can exhibit systematic biases and heterogeneous performance across subgroups, which may translate into inequities in downstream decisions. Importantly, fairness is not a purely statistical property of a model: its consequences depend on the societal and operational context in which predictions are used. For example, systematically higher risk scores for one group may intensify punitive actions in recidivism settings, but may also increase access to beneficial interventions in health-care triage. Therefore, fairness assessments should be interpreted in light of the decision rule and the relative costs of errors across groups.

The proposed cross-group concordance metric, xCI, is designed for time-to-event settings and summarizes both within-group and between-group ranking performance. Whereas the traditional CI primarily reflects within-group discrimination, xCI directly evaluates how a model orders individuals across groups among comparable pairs. Although both xCI and concordance imparity are built from group-conditional concordance quantities, they address different questions and can be complementary in practice. Concordance imparity, defined as the maximum difference between group-conditional CIs, is most informative for auditing discrimination gaps: it answers whether the model separates risk equally well within each group. In contrast, xCI is directional. Specifically, $xCI_{a,b}$ measures the probability that an individual from group $a$ who experiences the event earlier is ranked ahead of an individual from group $b$ who experiences the event later, while $xCI_{b,a}$ evaluates the reverse ordering. Comparing $xCI_{a,b}$ and $xCI_{b,a}$ therefore reveals cross-group ranking asymmetries, indicating which group is systematically prioritized in pairwise comparisons.

This directional cross-group perspective is particularly relevant when decisions depend on relative prioritization under censoring, such as referral queues, waitlists, or rationing limited resources. In these settings, relative ranking can be the primary driver of who receives an intervention first, making cross-group ordering an essential component of fairness evaluation. Conversely, when interventions are broadly available and decisions are not competitive (e.g., low-cost preventive services offered to all eligible individuals), cross-group ranking asymmetries may be less consequential than other properties such as calibration, threshold-based error rates, or subgroup-specific clinical utility.

Several limitations merit consideration. First, the number of xCI quantities grows with the number of groups, which can complicate reporting and interpretation; summary strategies or pre-specified contrasts may be needed in multi-group settings. Second, computation can be demanding because xCI requires identifying comparable pairs both within and across groups, motivating efficient implementations for large-scale data. Third, as with concordance-based estimands more generally, xCI is defined through comparable pairs and may be sensitive to censoring patterns that affect comparability, underscoring the importance of the censoring assumptions articulated in the Methods. Finally, while xCI provides a principled way to characterize cross-group ranking asymmetry, the normative question of how such asymmetries should be weighted against other fairness and utility criteria is context dependent and beyond the scope of this work.

\section{Conclusion}
In conclusion, we proposed a novel fairness metric based on the CI, which reveals ranking accuracy across different predefined groups. Compared to existing metrics, xCI offers a straightforward interpretation and a unique comparison perspective in the time-to-event setting. We provided a consistent estimator for this metric and demonstrated the relationship between the predicted risk scores and xCI. Additionally, we showed the application and interpretation of our metric using a publicly available dataset and large private clinical datasets for models we build and existing risk prediction equations. Finally, we offered an in-depth discussion on how this metric contributes to algorithmic fairness. We encourage researchers and practitioners to use xCI alongside CI for a more comprehensive assessment of fairness, particularly in the context of separation.

\section{Acknowledgment}
\begin{itemize}
 \item This study was funded by the Doris Duke Foundation in a grant awarded to the American Heart Association.
\end{itemize}
\bibliography{refs}

\begin{appendices}
   \section{Proofs for Section~\ref{cross:concordance}}\label{appendix:weighted average}
\begin{proof}[Proof of Proposition~\ref{CI:weighted:prop}]
    We consider the relationship between the proposed estimators of the xCI and CI. We define $\mathcal{S}$ and $\mathcal{S}^c$ as the following:

\begin{equation*}
|\mathcal{\mathcal{S}}| = \sum_{a \in \mathcal{G}} \sum_{b \in \mathcal{G}} |\mathcal{\mathcal{S}}_{a,b}|
\end{equation*}

\begin{equation*}
|\mathcal{\mathcal{S}}^c| = \sum_{a \in \mathcal{G}} \sum_{b \in \mathcal{G}} |\mathcal{\mathcal{S}}_{a,b}^c|
\end{equation*}

 The definition of  $\mathcal{S}_{a,b}$ ensures its non-overlapping nature: $\mathcal{S}_{a,b} \cap \mathcal{S}_{a',b'} = \emptyset \quad \text{unless} \ a = a' \ \text{and} \ b = b'$. Thus $|\mathcal{\mathcal{S}}|$ represents total number of comparable pairs and $|\mathcal{\mathcal{S}}^c|$ represents total number of concordant pairs. 
 
Then, $\hat{\text{CI}}$ is a weighted average of the $\hat{\text{xCI}}$:

\begin{align*}
    \hat{CI} &= |\mathcal{\mathcal{S}}^c|/|\mathcal{\mathcal{S}}|\\ \notag
    &= \frac{\sum_{a \in \mathcal{G}} \sum_{b \in \mathcal{G}} |\mathcal{\mathcal{S}}_{a,b}^c|}{|\mathcal{\mathcal{S}}|}\\ \notag
    &= \frac{\sum_{a \in \mathcal{G}} \sum_{b \in \mathcal{G}} \big(|\mathcal{\mathcal{S}}_{a,b}| * \hat{\text{xCI}}_{a,b}\big)}{|\mathcal{\mathcal{S}}|}\\
    &= \sum_{a \in \mathcal{G}} \sum_{b \in \mathcal{G}} \big(w_{a,b} * \hat{\text{xCI}}_{a,b}\big) \notag
\end{align*}

where $w_{a,b} = |\mathcal{\mathcal{S}}_{a,b}|/|\mathcal{\mathcal{S}}|$.
\end{proof}

\section{Proof of Proposition~\ref{prop:ipcw_consistency_main}}
\label{app:ipcw_proof}

We prove consistency of \eqref{eq:ipcw_xci_est} in two steps:
(i) consistency of an \emph{oracle} version using the true censoring survival functions;
(ii) a plug-in argument showing $\widehat K_g$ can replace $K_g$ without changing the limit.

\subsection{Oracle IPCW estimator}
Define the oracle weight
\[
w_{a,b}(t) := K_a(t)^{-1}K_b(t)^{-1},
\]
and the oracle numerator and denominator
\begin{align}
\label{eq:oracle_num}
N_n^\star
&:= \frac{1}{n_a n_b}
\sum_{i:\,G_i=a}\ \sum_{\substack{j:\,G_j=b\\ j\neq i}}
\Delta_i\, w_{a,b}(O_i)\, I(O_i<O_j,\; O_i<\tau)\, I(R_i>R_j),\\
\label{eq:oracle_den}
D_n^\star
&:= \frac{1}{n_a n_b}
\sum_{i:\,G_i=a}\ \sum_{\substack{j:\,G_j=b\\ j\neq i}}
\Delta_i\, w_{a,b}(O_i)\, I(O_i<O_j,\; O_i<\tau),
\end{align}
where $n_a:=\sum_{i=1}^n I(G_i=a)$ and $n_b:=\sum_{i=1}^n I(G_i=b)$.

\begin{lemma}[Limits of oracle numerator/denominator]
\label{lem:oracle_limits}
Under Assumptions 1--2,
\begin{align*}
D_n^\star &\xrightarrow{P} \Pr(T_i<T_j,\; T_i<\tau,\; G_i=a,\; G_j=b),\\
N_n^\star &\xrightarrow{P} \Pr(R_i>R_j,\; T_i<T_j,\; T_i<\tau,\; G_i=a,\; G_j=b).
\end{align*}
\end{lemma}

\begin{proof}
We prove the denominator; the numerator is analogous with an additional indicator $I(R_i>R_j)$.

Consider the kernel
\[
h^\star((T_i,U_i,G_i,R_i),(T_j,U_j,G_j,R_j))
:= \Delta_i\, w_{a,b}(O_i)\, I(O_i<O_j,\; O_i<\tau)\, I(G_i=a,G_j=b).
\]
By Assumption 2, $K_a(t)\ge \kappa$ and $K_b(t)\ge \kappa$ on $[0,\tau]$, hence
\[
0 \le h^\star(\cdot,\cdot) \le \kappa^{-2},
\]
so the kernel is bounded. Since the double sum in \eqref{eq:oracle_den} is a two-sample (generalized) U-statistic
with bounded kernel, a law of large numbers for generalized/two-sample U-statistics ~\citep{kowalski2008modern} yields
\[
D_n^\star \xrightarrow{P} \mathbb E\!\left[h^\star((T_i,U_i,G_i,R_i),(T_j,U_j,G_j,R_j))\right],
\]
where $(i,j)$ are independent draws from the population.

Next, simplify the expectation. Note that if $\Delta_i=1$ then $O_i=T_i$, so
\[
\Delta_i I(O_i<O_j,\; O_i<\tau)
= I(U_i\ge T_i)\, I\!\left(T_i<\min(T_j,U_j),\; T_i<\tau\right).
\]
Because $T_i<\min(T_j,U_j)$ is equivalent to $T_i<T_j$ and $T_i<U_j$, we have
\[
\Delta_i I(O_i<O_j,\; O_i<\tau)
= I(T_i<T_j,\; T_i<\tau)\, I(U_i\ge T_i)\, I(U_j>T_i).
\]
Therefore,
\begin{align*}
&\mathbb E\!\left[\Delta_i\, w_{a,b}(O_i)\, I(O_i<O_j,\; O_i<\tau)\, I(G_i=a,G_j=b)\right]\\
&\qquad=
\mathbb E\!\left[
w_{a,b}(T_i)\, I(T_i<T_j,\; T_i<\tau)\, I(G_i=a,G_j=b)\,
I(U_i\ge T_i)\, I(U_j>T_i)
\right].
\end{align*}
Using independence across individuals and Assumption 2 (non-informative censoring within groups),
\[
\mathbb E[I(U_i\ge T_i)\mid T_i, G_i=a]=K_a(T_i), \qquad
\mathbb E[I(U_j>T_i)\mid T_i, G_j=b]=K_b(T_i).
\]
Hence the weight cancels:
\[
w_{a,b}(T_i)\,K_a(T_i)\,K_b(T_i)=1,
\]
and the expectation reduces to
\[
\mathbb E\!\left[I(T_i<T_j,\; T_i<\tau)\, I(G_i=a,G_j=b)\right]
=\Pr(T_i<T_j,\; T_i<\tau,\; G_i=a,\; G_j=b).
\]
This proves the denominator limit. The numerator limit follows identically with the extra factor $I(R_i>R_j)$.
\end{proof}

\subsection{Plug-in censoring weights}
Define the plug-in numerator/denominator (using $\widehat w_{a,b}$ from \eqref{eq:ipcw_weight})
\begin{align*}
N_n &:=
\frac{1}{n_a n_b}
\sum_{i:\,G_i=a}\ \sum_{\substack{j:\,G_j=b\\ j\neq i}}
\Delta_i\, \widehat w_{a,b}(O_i)\, I(O_i<O_j,\; O_i<\tau)\, I(R_i>R_j),\\
D_n &:=
\frac{1}{n_a n_b}
\sum_{i:\,G_i=a}\ \sum_{\substack{j:\,G_j=b\\ j\neq i}}
\Delta_i\, \widehat w_{a,b}(O_i)\, I(O_i<O_j,\; O_i<\tau),
\end{align*}
so that $\widehat{\mathrm{xCI}}_{a,b,\tau} = N_n/D_n$.

\begin{lemma}[Plug-in error is asymptotically negligible]
\label{lem:plugin_negligible}
Under Assumptions 1--3,
\[
N_n - N_n^\star = o_p(1), \qquad D_n - D_n^\star = o_p(1),
\]
where $N_n^\star,D_n^\star$ are the oracle quantities in \eqref{eq:oracle_num}--\eqref{eq:oracle_den}.
\end{lemma}

\begin{proof}
We prove the statement for the denominator; the numerator is analogous.

By Assumption 2, $K_g(t)\ge \kappa$ on $[0,\tau]$ and by Assumption 3, $\widehat K_g$ is uniformly consistent on $[0,\tau]$.
It follows (by continuity of $x\mapsto 1/x$ on $[\kappa/2,\infty)$ and standard arguments) that
\begin{equation}
\label{eq:weight_uniform}
\sup_{t\in[0,\tau]}\left|\widehat w_{a,b}(t)-w_{a,b}(t)\right| = o_p(1).
\end{equation}
Then
\begin{align*}
|D_n - D_n^\star|
&\le
\sup_{t\in[0,\tau]}\left|\widehat w_{a,b}(t)-w_{a,b}(t)\right|\cdot
\frac{1}{n_a n_b}\sum_{i:\,G_i=a}\sum_{j:\,G_j=b}
\Delta_i\, I(O_i<O_j,\; O_i<\tau).
\end{align*}
The double-sum average is bounded by 1, hence the right-hand side is $o_p(1)$ by \eqref{eq:weight_uniform}.
\end{proof}

\subsection{consistency of $\widehat{\mathrm{xCI}}_{a,b,\tau}$}
\begin{proof}[Proof of Proposition~\ref{prop:ipcw_consistency_main}]
By Lemma~\ref{lem:oracle_limits}, $(N_n^\star,D_n^\star)$ converges in probability to
\[
(N,D)
=
\Bigl(
\Pr(R_i>R_j,\; T_i<T_j,\; T_i<\tau,\; G_i=a,\; G_j=b),\;
\Pr(T_i<T_j,\; T_i<\tau,\; G_i=a,\; G_j=b)
\Bigr).
\]
By Lemma~\ref{lem:plugin_negligible}, $(N_n,D_n)=(N_n^\star,D_n^\star)+o_p(1)$, hence
\[
(N_n,D_n)\xrightarrow{P}(N,D).
\]
Under Assumption 1, the conditioning event has positive probability, so $D>0$.
Therefore, by Slutsky's theorem~\citep{van2000asymptotic} and continuity of $(x,y)\mapsto x/y$ at $(N,D)$,
\[
\widehat{\mathrm{xCI}}_{a,b,\tau}=\frac{N_n}{D_n}\xrightarrow{P}\frac{N}{D}
=\Pr(R_i>R_j \mid T_i<T_j,\; T_i<\tau,\; G_i=a,\; G_j=b)
=\mathrm{xCI}_{a,b,\tau}(R).
\]
\end{proof}
\subsection{Optional: estimated risk score $R_i=\widehat\beta^\top X_i$}
If the risk score is obtained from an estimated parameter $\widehat\beta$,
then Proposition~\ref{prop:ipcw_consistency_main} yields consistency for the xCI corresponding to the
random score $R_i=\widehat\beta^\top X_i$. To identify a deterministic limit, impose the following:

\begin{enumerate}
\setcounter{enumi}{3}
\item (Risk-score convergence and no-ties condition).
$\widehat\beta \xrightarrow{P}\beta_0$ and
\[
\Pr\!\left(\beta_0^\top(X_i-X_j)=0 \,\middle|\, T_i<T_j,\; T_i<\tau,\; G_i=a,\; G_j=b\right)=0.
\]
\end{enumerate}

Define
\[
\mathrm{xCI}_{a,b,\tau}(\beta_0)
:= \Pr\!\left(\beta_0^\top X_i > \beta_0^\top X_j \,\middle|\, T_i<T_j,\; T_i<\tau,\; G_i=a,\; G_j=b\right).
\]

\begin{prop}[Limit when $R_i=\widehat\beta^\top X_i$]
\label{prop:beta_limit}
Under Assumptions 1--3 and Assumption 4, if $R_i=\widehat\beta^\top X_i$ then
\[
\widehat{\mathrm{xCI}}_{a,b,\tau}\xrightarrow{P}\mathrm{xCI}_{a,b,\tau}(\beta_0).
\]
\end{prop}

\begin{proof}
From Proposition~\ref{prop:ipcw_consistency_main},
\[
\widehat{\mathrm{xCI}}_{a,b,\tau}\xrightarrow{P}\mathrm{xCI}_{a,b,\tau}(\widehat\beta),
\quad\text{where}\quad
\mathrm{xCI}_{a,b,\tau}(\widehat\beta):=
\Pr\!\left(\widehat\beta^\top X_i > \widehat\beta^\top X_j \mid T_i<T_j,\; T_i<\tau,\; G_i=a,\; G_j=b\right).
\]
It remains to show $\mathrm{xCI}_{a,b,\tau}(\widehat\beta)\xrightarrow{P}\mathrm{xCI}_{a,b,\tau}(\beta_0)$.
Let $\Delta X:=X_i-X_j$. On the event $\{\widehat\beta\neq \beta_0\}$,
\[
\bigl\{ \widehat\beta^\top \Delta X > 0,\; \beta_0^\top \Delta X \le 0 \bigr\}
\subseteq
\bigl\{ |\beta_0^\top \Delta X| \le \|\widehat\beta-\beta_0\|\,\|\Delta X\| \bigr\},
\]
and similarly with the roles reversed. By Assumption 4, $\|\widehat\beta-\beta_0\|\xrightarrow{P}0$,
and the no-ties condition ensures $\Pr(\beta_0^\top \Delta X=0 \mid \cdot)=0$, implying
\[
I(\widehat\beta^\top \Delta X>0)\xrightarrow{P} I(\beta_0^\top \Delta X>0)
\quad\text{for a random comparable pair.}
\]
Dominated convergence then yields
$\mathrm{xCI}_{a,b,\tau}(\widehat\beta)\to \mathrm{xCI}_{a,b,\tau}(\beta_0)$ in probability, and the result follows by Slutsky~\citep{van2000asymptotic}.
\end{proof}

\section{Proof of Proposition~\ref{prop:impact_predicted_group_diff}}
\label{appendix:impact_predicted_group_diff}

Consider a pair $(i,j)$ with $G_i=1$ and $G_j=0$. Write
\[
\Delta_{ij} Z:=Z_i-Z_j\sim\mathcal N(0,2\sigma_z^2),\qquad
\Delta\epsilon:=\epsilon_j-\epsilon_i\sim\mathcal N(0,2\sigma^2),
\]
independent of each other. Define
\[
D:=\beta_1+\beta_2\Delta_{ij} Z-\Delta\epsilon,\qquad
\sigma_D^2:=\Var(D)=2(\beta_2^2\sigma_z^2+\sigma^2).
\]
Then $(\Delta_{ij} Z,D)$ is bivariate normal with mean $(0,\beta_1)$ and
\[
\Cov(\Delta_{ij} Z,D)=2\beta_2\sigma_z^2,\qquad
\Var(\Delta_{ij} Z)=2\sigma_z^2,\qquad
\Var(D)=\sigma_D^2.
\]
The true ordering $T_i<T_j$ is equivalent to $D<0$.

Let
\[
\alpha:=\frac{0-\\E[D]}{\sigma_D}=-\frac{\beta_1}{\sigma_D},\qquad
\lambda(\alpha):=\frac{\phi(\alpha)}{\Phi(\alpha)},\qquad
\delta(\alpha):=\lambda(\alpha)\bigl(\lambda(\alpha)-\alpha\bigr),
\]
and for right truncation define
\[
\lambda_R(\alpha):=\frac{\phi(\alpha)}{1-\Phi(\alpha)},\qquad
\delta_R(\alpha):=\lambda_R(\alpha)\bigl(\lambda_R(\alpha)+\alpha\bigr),
\]
where $\phi,\Phi$ are the standard normal probability density function and cumulative density function. For a scalar $X\sim\mathcal N(\mu,\sigma^2)$,
\[
\E[X\mid X<0]=\mu-\sigma\,\lambda\!\left(\tfrac{0-\mu}{\sigma}\right),
\quad
\Var(X\mid X<0)=\sigma^2\Bigl(1-\delta\!\left(\tfrac{0-\mu}{\sigma}\right)\Bigr),
\]
and similarly with $\lambda_R,\delta_R$ for $X\mid X>0$~\citep{tallis1961moment, horrace2005some}.

Using $\E[\Delta_{ij} Z\mid D=d]=\frac{\Cov(\Delta_{ij} Z,D)}{\Var(D)}(d-\E[D])$ and the laws of total expectation/variance:

For Left truncation ($D<0$, i.e., $a$ vs.\ $b$):
\[
\mu_c:=\E[\Delta_{ij} Z\mid D<0]
=-\,\frac{2\beta_2\sigma_z^2}{\sigma_D}\,\lambda(\alpha),
\]
\[
\sigma_c^2:=\Var(\Delta_{ij} Z\mid D<0)
=2\sigma_z^2-\frac{(2\beta_2\sigma_z^2)^2}{\sigma_D^2}\,\delta(\alpha).
\]

The score difference is $\Delta_{i,j} \hat r:=\hat r_i-\hat r_j=\hat\beta_1+\hat\beta_2\Delta_{ij} Z$.

For $a$ vs.\ $b$ ($D<0$).
Given $D<0$,
\[
\Delta_{i,j} \hat r\;\sim\;\mathcal N\!\big(\hat\beta_1+\hat\beta_2\mu_c,\;\hat\beta_2^2\sigma_c^2\big),
\]
since $(\Delta_{ij} Z, D)$ is jointly normal and $\Delta_{i,j}\hat r$ is an affine function of $\Delta_{ij} Z$, which preserves normality and linearly maps the conditional mean/variance.
So
\[
\text{xCI}_{a,b}
=\Pr(\Delta\hat r>0\mid D<0)
=\Phi\!\left(
\frac{\hat\beta_1+\hat\beta_2\,\mu_c}{\sqrt{\hat\beta_2^2\,\sigma_c^2}}
\right).
\]

For right truncation ($D>0$, i.e., $b$ vs.\ $a$).
\[
\mu_c':=\E[\Delta_{ij} Z\mid D>0]
=\frac{2\beta_2\sigma_z^2}{\sigma_D}\,\lambda_R(\alpha),
\]
\[
\sigma_c'^2:=\Var(\Delta_{ij} Z\mid D>0)
=2\sigma_z^2-\frac{(2\beta_2\sigma_z^2)^2}{\sigma_D^2}\,\delta_R(\alpha).
\]

Given $D>0$,
\[
\Delta_{i,j}\hat r\;\sim\;\mathcal N\!\big(\hat\beta_1+\hat\beta_2\mu_c',\;\hat\beta_2^2\sigma_c'^2\big),
\]
by the same joint-normality and affine-transformation argument as above; the correct ranking event is $\Delta_{i,j} \hat r<0$, hence
\[
\text{xCI}_{b,a}
=\Pr(\Delta_{i,j} \hat r<0\mid D>0)
=\Phi\!\left(
-\frac{\hat\beta_1+\hat\beta_2\,\mu_c'}{\sqrt{\hat\beta_2^2\,\sigma_c'^2}}
\right).
\]

When $G_i=G_j$, $\Delta_{i,j} \hat r=\hat\beta_2\Delta_{ij} Z$ and $D_{w}:=\beta_2\Delta_{ij} Z-\Delta\epsilon$ has mean $0$ and variance $2(\beta_2^2\sigma_z^2+\sigma^2)$.
With $\alpha=0$, $\lambda(0)=\sqrt{2/\pi}$ and $\delta(0)=2/\pi$ yield
\[
\mu_{w}:=\E[\Delta_{ij} Z\mid D_{w}<0]
=-\,\frac{2\beta_2\sigma_z^2}{\sigma_D}\,\sqrt{\frac{2}{\pi}},\qquad
\sigma_{w}^2:=\Var(\Delta_{ij} Z\mid D_{w}<0)
=2\sigma_z^2-\frac{(2\beta_2\sigma_z^2)^2}{\sigma_D^2}\,\frac{2}{\pi}.
\]
Therefore,
\[
\text{xCI}_{a,a}=\text{xCI}_{b,b}
=\Phi\!\left(\frac{\hat\beta_2\,\mu_{w}}{\sqrt{\hat\beta_2^2\,\sigma_{w}^2}}\right).
\]

\end{appendices}
\end{document}